\documentclass[twocolumn,showpacs,amsfonts,aps,prl,floatfix,
nofootinbib,float]{revtex4} 

\usepackage{amsmath}
\usepackage{bm}
\usepackage{graphicx}

\newcommand{\be}[1]{\begin{equation}\label{#1}}
\newcommand{\ee}{\end{equation}}

\usepackage{color}

\begin{document}

\title{Electron-muon correlation as a new probe to strongly interacting quark-gluon plasma}
\date{\today}

\author{Yukinao Akamatsu}
\affiliation{Department of Physics, The University of Tokyo, Tokyo 113-0033, Japan}

\author{Tetsuo Hatsuda}
\affiliation{Department of Physics, The University of Tokyo, Tokyo 113-0033, Japan}

\author{Tetsufumi Hirano}
\affiliation{Department of Physics, The University of Tokyo, Tokyo 113-0033, Japan}

\begin{abstract}
As a new and clean probe to the strongly interacting quark-gluon plasma (sQGP), 
we propose  an azimuthal correlation of an electron and a muon which originate from the semileptonic decay of charm and bottom quarks. By solving the Langevin equation for the 
 heavy quarks under the  hydrodynamic evolution of the hot plasma, we  
 show that substantial  quenching of the away-side peak in
  the electron-muon correlation can be seen if the sQGP drag force acting on heavy quarks 
   is large enough as suggested  from the gauge/gravity correspondence.
  The effect could be detected in high-energy heavy-ion collisions
at the Relativistic Heavy Ion Collider and the Large Hadron Collider.
\end{abstract}


\maketitle
Quark-gluon plasma (QGP) is a novel state of matter expected to exist at extremely high temperature in early universe. 
Laboratory studies of the QGP by means of the high energy nuclear collisions 
are underway at the Relativistic Heavy Ion Collider (RHIC) and will be continued at the Large Hadron Collider (LHC) \cite{QGP}.
The space-time evolution of the QGP created at RHIC is successfully described by ideal hydrodynamics where viscous effects are assumed to be small \cite{Hirano:2008hy}.
This indicates that the quarks and gluons in the QGP are strongly interacting even above the critical temperature of the deconfinement transition in accordance with previous theoretical expectations \cite{Linde:1980ts}.
The small value of a ratio of the shear viscosity $\eta$ to the entropy density $s$ recently obtained from the gauge/gravity correspondence \cite{Son:2007vk}
and from the lattice quantum chromodynamics (QCD) simulations \cite{Nakamura:2004sy} strengthens the idea of the strongly interacting QGP, or the sQGP in short. 

The hard probes such as high energy jets, heavy quarks, and quarkonia created in the initial stage of the heavy ion collisions would provide us with  experimental information on QGP.
For example, suppression of the jet events and the disappearance of the jet correlations are intimately related to the parton energy loss inside QGP \cite{Bjorken:1982tu,Gyulassy:1990ye}.  
Suppression of the dileptons from the electromagnetic decay of the heavy quarkonia is related to the screening of the heavy quark potentials in QGP \cite{Matsui:1986dk}.
Semileptonic decays of charm $c$ and bottom $b$ also carry information on the interaction of heavy quarks with QGP \cite{Rapp}:
Recent data of the energetic single electrons from heavy quark decays \cite{phenix2007,star2007} indicate a rather large energy loss of $c$ and $b$ inside the hot medium \cite{akamatsu,Moore:2004tg,vanHees:2004gq,Armesto:2005mz,Wicks2007,Gossiaux:2008jv}.
 
The purpose of this Letter is to propose a new observable which further signifies the dynamical property of heavy quarks in QGP: 
It is an azimuthal correlation of an electron ($e$) and a muon ($\mu$) which originate from the semileptonic decays of back-to-back heavy quarks traversing the plasma. 
In particular, quenching of the  away-side peak in the $e$-$\mu$ correlation turns out to be a clean observable to extract the in-medium energy-loss of heavy quarks \cite{akamatsu_QM09}. 
There are several advantages to look at the $e$-$\mu$ correlation over the other correlations: 
First of all, leptons from semileptonic decays keep the direction of their parent heavy quarks as long as their transverse momenta are large enough (e.g. $p_{T} > 3$ GeV/$c$ for electrons) \cite{wang}.
Therefore, the dileptons reflect the correlation of a heavy quark--anti-quark pair just prior to its semileptonic decay.
Furthermore, the $e$-$\mu$ pair does not receive contamination from virtual photons and neutral vector mesons, so that the backgrounds are significantly reduced in comparison to the $e^+$-$e^-$ and $\mu^+$-$\mu^-$ correlations.
The lepton-hadron azimuthal correlation \cite{star_new,machcone} could also provide us with useful information about the heavy quarks in the medium. 
However, the signal may be contaminated by other hadronic sources and interactions.
Because of these reasons, the $e$-$\mu$ azimuthal correlation would be an ideal probe to study the heavy quark propagation in the sQGP.

In the following, we describe the dynamics of heavy quarks  by the relativistic Langevin equation with hydrodynamic background, which is an approach recently 
developed by the present authors \cite{akamatsu}.
The hot matter created in the heavy ion collisions at RHIC is treated as a perfect fluid obeying the relativistic hydrodynamics equations \cite{Hirano_Tsuda}:  
\begin{eqnarray}
\label{eq:hydro}
&&\partial_{\mu}T^{\mu\nu}=0, \\
&&T^{\mu\nu}=(\varepsilon +P)u^{\mu}u^{\nu}-Pg^{\mu\nu},
\end{eqnarray}
where $T^{\mu \nu}$ is the energy-momentum tensor, $\varepsilon$ is the local energy density, $P$ is the local pressure, and $u^{\mu}$ is the local flow velocity.
Initial conditions of the space-time evolution are determined to reproduce 
the experimental hadronic observables in low $p_{T}$ regions. 

The heavy quark dynamics in the hot matter is described by the relativistic Langevin equation formulated in the rest frame of a fluid element \cite{akamatsu,Debbasch},
\begin{eqnarray}
\label{eq:langevin1}
&&\Delta \vec x = \frac{\vec p}{E(p)}\Delta t, \ \ 
\Delta\vec p =-\Gamma(p)\vec p\Delta t +\vec \xi(t),\\
\label{eq:langevin2}
&&\langle\xi _{i}(t)\xi _{j}(t')\rangle =D_{ij}(p)\delta _{tt'}\Delta t,
\end{eqnarray}
where $\vec x$ and $\vec p$ are the position and the momentum of a heavy quark with mass $M$, $E(p)=\sqrt{{\vec p}^2 + M^2}$ is the kinetic energy of a heavy quark, $\Gamma(p)$ is the drag coefficient, and $\vec \xi(t)$ is the Gaussian white noise with $D_{ij}(p)$ being the momentum dependent diffusion constant. 
Following our previous study, we adopt a particular parametrization,
 $\Gamma(p)=\gamma T^2/M$, where $T$ is local temperature and $\gamma$ is
a drag parameter: 
This is motivated by  the drag force obtained by using the gauge/gravity correspondence \cite{AdS_Drag}.
Then, the relativistic fluctuation-dissipation theorem leads to
$D_{ij}(p)=2\gamma T^3 (E+T)\delta _{ij}/M$ \cite{akamatsu}.
More sophisticated form with an anisotropic diffusion constant would be necessary in the future quantitative studies. 
   
Heavy quarks are produced through the initial hard processes simulated by the Monte Carlo event generator PYTHIA 6.4 \cite{PYTHIA}. After the production, 
 they are diffused inside the QGP fluid according to Eqs.~(\ref{eq:langevin1}) and (\ref{eq:langevin2}) until the surrounding temperature decreases and reaches the critical temperature $T_c = 170$ MeV.
Below $T_c$, heavy quarks hadronize to $D$ or $B$ mesons which subsequently undergo semileptonic decays simulated by PYTHIA.
The color-singlet $D$ and $B$ mesons are assumed to propagate along the free-streaming path in the confined hadronic matter.  

Analysis of the nuclear modification factor $R_{AA}$ for energetic single electrons at RHIC 
 on the basis of the Langevin + hydro approach leads to
the drag parameter $\gamma$ = 1--3 for heavy quarks \cite{akamatsu}.
This value is consistent with  $\gamma \sim 2$ estimated from the gauge/gravity 
correspondence applied to hot QCD matter \cite{Gubser:2006qh}, while it is much larger than $\gamma \sim 0.2$ given by the leading order perturbative QCD calculation \cite{Moore:2004tg}.
If the heavy quarks were completely thermalized, the correlation of a heavy quark--anti-quark pair would be washed out.
However, this is not the case as shown in our previous study \cite{akamatsu}:
Using the averaged temperature $\langle T\rangle \sim 0.21$ GeV for the charm and bottom quarks in hot matter created by collisions at RHIC, one can estimate their relaxation time $\tau_Q = M_Q/\gamma \langle T\rangle^2$.
This together with the averaged staying time (3--4 fm/$c$) of charm and bottom quarks inside the QGP fluid suggests that the charm quarks are partially thermalized while bottom quarks are not thermalized at RHIC for $\gamma$ = 1--3  \cite{akamatsu}.
Therefore the correlation of a heavy quark--anti-quark pair is expected to survive partially during their propagation and serve as a good probe of the drag coefficient of heavy quarks.

\begin{figure}[h]
\centering
\vspace{0.2cm}
\includegraphics[width=7.5cm,clip]{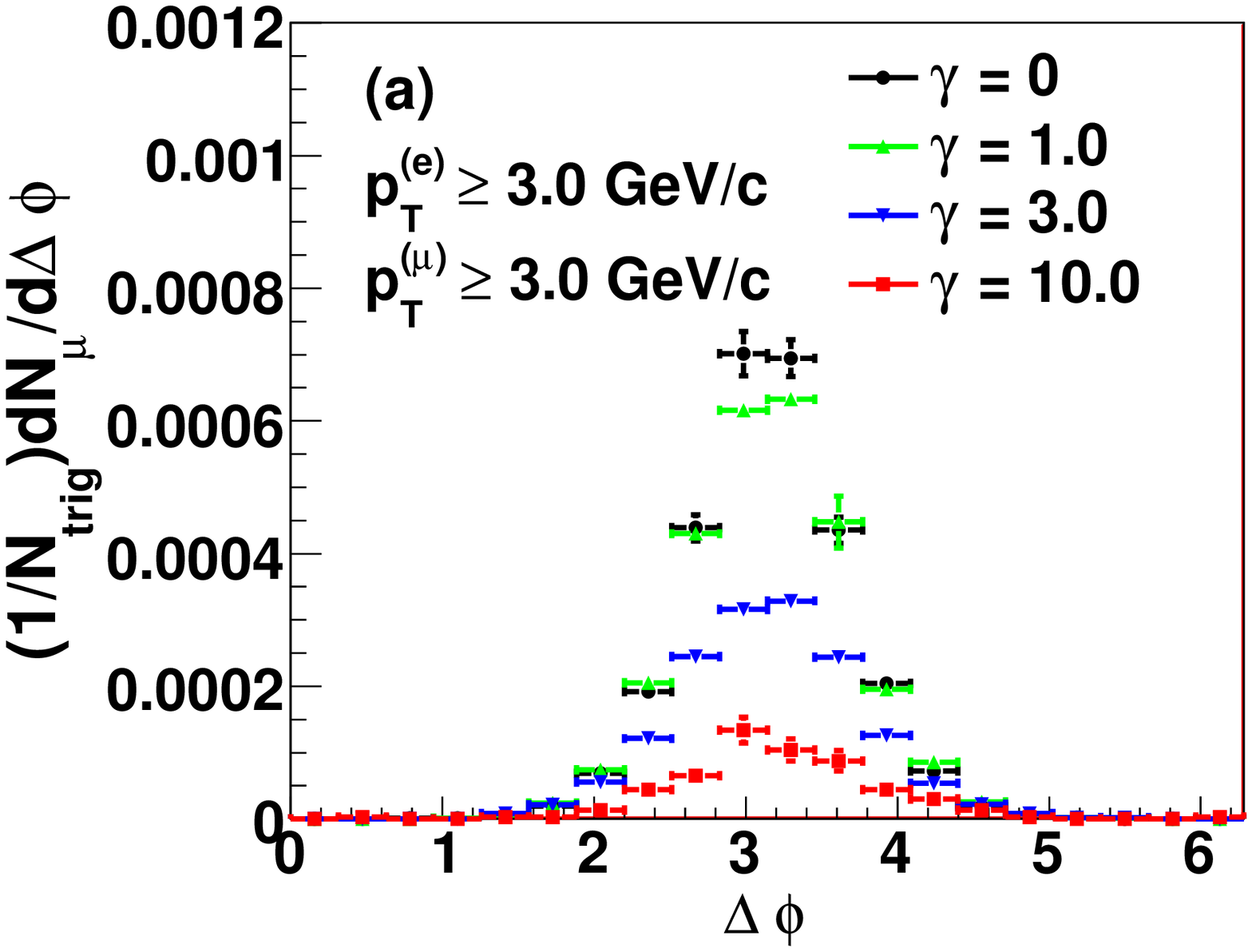}
\includegraphics[width=7.5cm,clip]{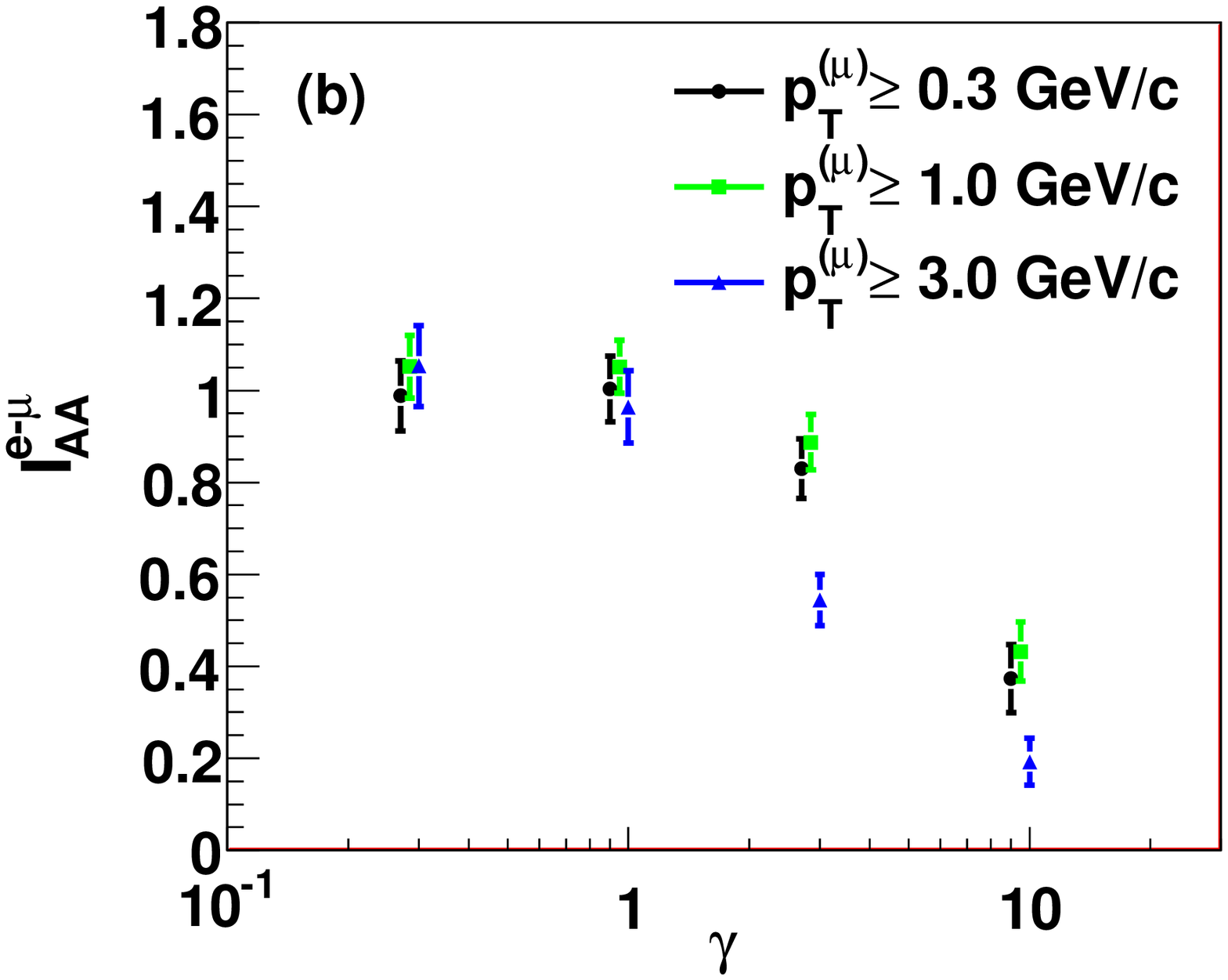}
\caption[e-m]
{
(a) Azimuthal correlation of electrons and muons decayed from a heavy quark--anti-quark pair.
Both the trigger electrons and the associate muons satisfy $p_{T}\geq 3.0$ GeV/$c$.
Trigger electrons are in mid-pseudorapidity $|\eta|\leq 0.35$ and associate muons
are in forward-pseudorapidity $1.4\leq|\eta|\leq 2.1$. 
(b) Away-side quenching factor $I_{AA}^{e\mbox{-}\mu}$ as a function of the drag parameter $\gamma$ with three cases of the transverse momentum cut ($p^{(\mu)}_{T} \geq$ 0.3, 1.0, and 3.0 GeV/$c$) for associate muons.
Trigger electrons satisfy $p^{(e)}_{T}\geq 3.0$ GeV/$c$.
Plots with the same drag parameter $\gamma$ are slightly shifted to avoid overlap.
}
\label{e-m}
\end{figure}
 
Let us now consider the  $e$-$\mu$ azimuthal correlation obtained from our Langevin + hydro approach.
 The electrons and muons are from the semileptonic decays: $D\rightarrow l$, $B\rightarrow l$,
and $B\rightarrow D\rightarrow l$, where $l$ stands for the lepton.
Taking into account the current setups of the PHENIX detector, we take 
electrons with the pseudorapidity $|\eta|\leq$ 0.35 and the transverse momentum $p^{(e)}_{T}\geq$ 3.0 GeV/$c$ as trigger particles, and muons with 1.4 $\leq |\eta|\leq$ 2.1 
and $p^{(\mu)}_{T}\geq$ 3.0 GeV/$c$ as associate particles.
Collision geometry is fixed to be nearly central with the impact parameter $b$ = 3.1 fm.

Shown in Fig.~\ref{e-m} (a) is the $e$-$\mu$ azimuthal correlation per trigger electron as a function of the relative opening azimuthal angle $\Delta \phi$. 
If there is no energy loss of heavy quarks ($\gamma=0$), high momentum muons appear in the away-side as a result of the back-to-back nature of the heavy quark production. 
It should be noted that the width of the peak in the $\gamma = 0$ case comes
dominantly from finite primordial transverse momentum ($\sim$ 1 GeV/$c$) of initial gluons inside a nucleon in the default parameter set of PYTHIA and that we do not consider nuclear broadening nor shadowing in the parton distribution of a nucleus. 
  There is only one peak in the away side,  simply because 
a single semileptonic decay does not produce more than one lepton in the same direction.
As the energy loss increases, transverse momentum of the associate muons also decreases and the total yield in the away side with $p^{(\mu)}_{T}\geq$ 3.0 GeV/$c$ decreases.
This is clearly seen for $\gamma = 3$ and $10$ in  Fig.~\ref{e-m} (a). 

To quantify the effect of the suppression in the away-side muons and to compare the result with the future experiments, we introduce a quenching factor $I_{AA}$ as follows:
\begin{eqnarray}
\label{eq:measure}
\Sigma_{AA}& = &\int_{\phi_{\rm min}}^{\phi_{\rm max}} d(\Delta\phi)
\left[ \frac{1}{N_{\rm trig}}\frac{dN_{\rm assoc}}{d\Delta \phi} \right]_{\rm ZYAM} , \nonumber \\
I_{AA} & = &{\Sigma_{AA}}/{\Sigma_{pp}},
\end{eqnarray}
where $N_{\mathrm{trig}}$  and $N_{\mathrm{assoc}}$ are the number of trigger particles (the electrons here) and the number of associate particles (the muons here).
ZYAM implies zero-yield-at-minimum where muons other than the ones from heavy quark decays are subtracted as backgrounds.
Theoretically, $\Sigma_{AA}$ is a function of the drag parameter $\gamma$, while
  $\Sigma_{pp}$ (the case for $pp$ collision) corresponds to $\gamma=0$.
Taking  $(\phi_{\rm min},\phi _{\rm max})=(0,2\pi)$ in Fig.~\ref{e-m} (a), the away-side quenching factor $I_{AA}^{e\mbox{-}\mu}$ with $p^{(\mu)}_{T}\geq$ 0.3, 1.0, and 3.0 GeV/$c$ are evaluated as shown in Fig.~\ref{e-m} (b).  
 We find that the quenching factor is sufficiently sensitive to $\gamma$:
  In the case of  high $p_{T}$ muons, $I_{AA}^{e\mbox{-}\mu}$ 
   is  as small as  0.55 for relatively  large drag parameter $\gamma = 3$.
 We thus conclude that $I_{AA}^{e\mbox{-}\mu}$ is 
 a useful observable which carries  information of the heavy quarks in sQGP.

\begin{figure}[h]
\centering
\includegraphics[width=7.5cm,clip]{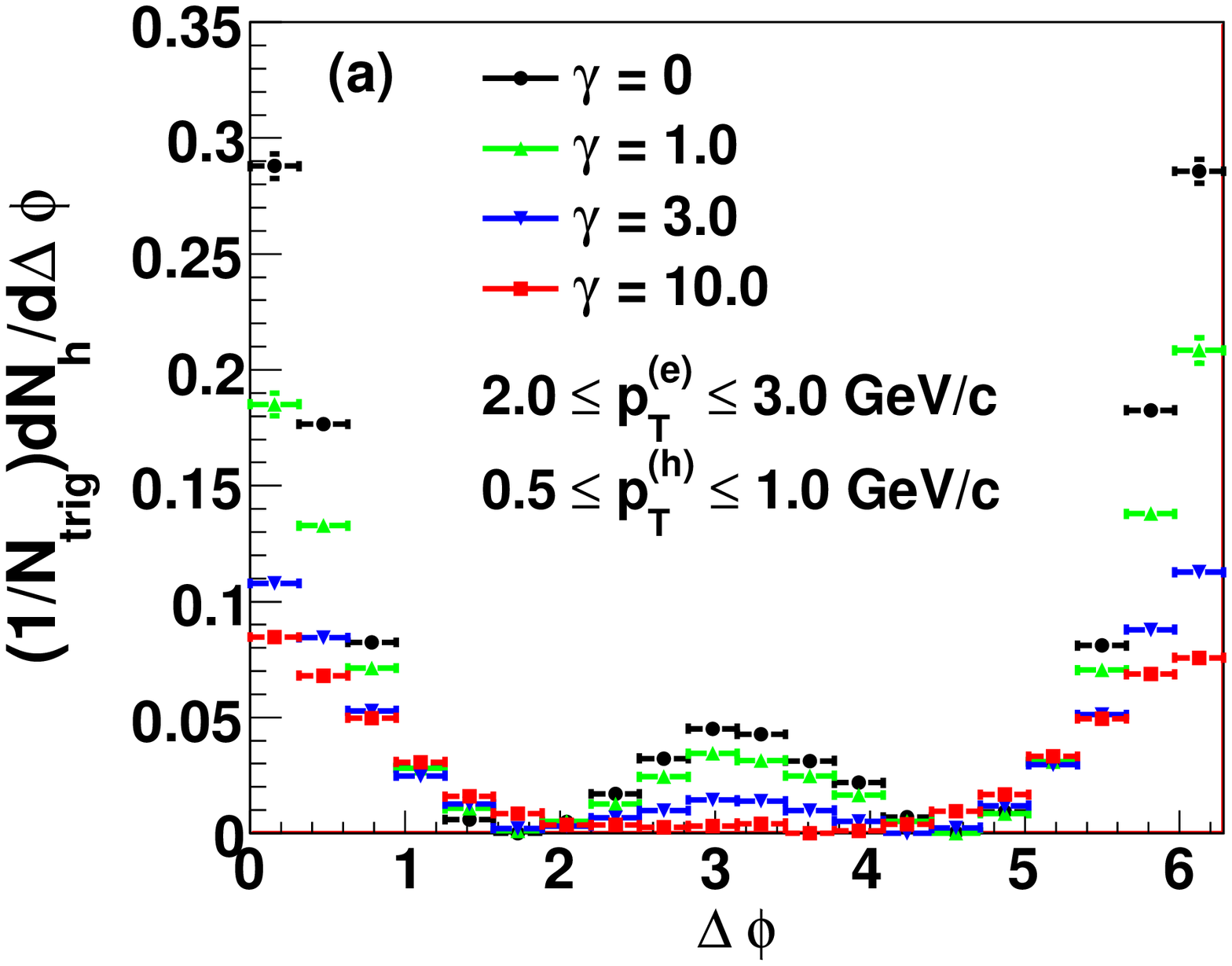}
\includegraphics[width=7.5cm,clip]{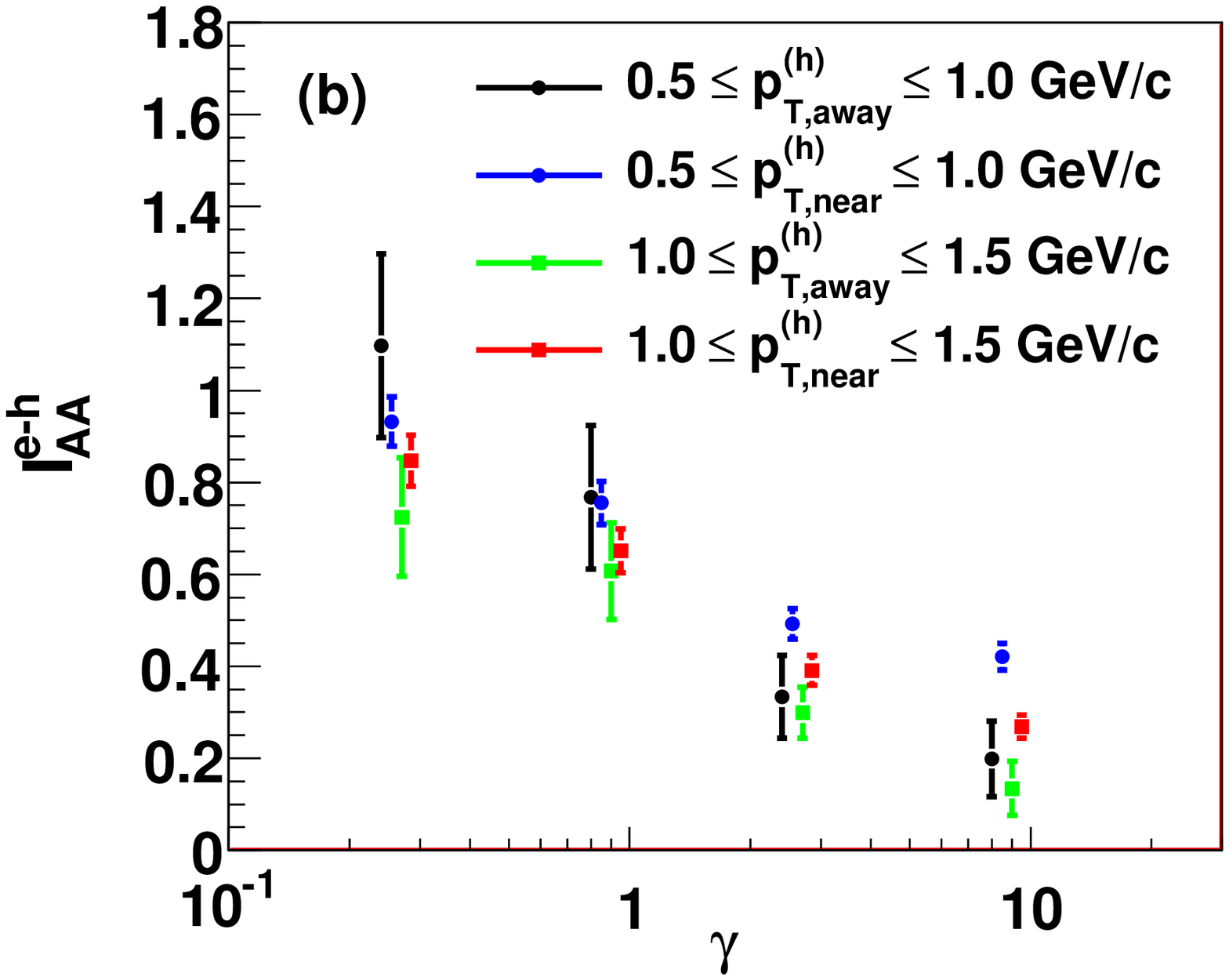}
\caption[R:e-h]
{
(a)
Azimuthal correlation of electrons and charged hadrons decayed from a heavy quark--anti-quark pair.
The trigger electrons satisfy 2.0 $\leq p_{T}^{(e)}\leq$ 3.0 GeV/$c$ and the associate charged hadrons satisfy 0.5 $\leq p_{T}^{(h)}\leq$ 1.0 GeV/$c$.
Both trigger and associate particles are in mid-pseudorapidity $|\eta|\leq$ 0.35.
(b) 
Near-side and away-side quenching factors $I_{AA}^{e\mbox{-}h}$ as a function of the drag parameter $\gamma$ with two cases of the transverse momentum window (0.5 $\leq p^{(h)}_T\leq$ 1.0 GeV/$c$ and 1.0 $\leq p^{(h)}_T\leq$ 1.5 GeV/$c$) for associate particles.
Trigger electrons  satisfy $2.0\leq p^{(e)}_{T}\leq 3.0$ GeV/$c$.
Plots with the same drag parameter $\gamma$ are slightly shifted to avoid overlap.
}
\label{e-h}
\end{figure}

Let us now consider the azimuthal correlation of an electron and a light charged hadron.
The latter includes the charged pions, the charged kaons and the protons from the decay of the $D$ and/or $B$-mesons.
Although the detection of the electron-hadron ($e$-$h$) correlation itself is less demanding statistically, there are several disadvantages of the $e$-$h$ correlation against the $e$-$\mu$ correlation:
(i) light hadrons interact strongly with the hadronic matter so that the azimuthal correlations are smeared by final state interactions, 
(ii) heavy quark propagation in the sQGP may affect the evolution of the 
flow of the hot matter and eventually introduces azimuthal angle dependence
 in  the production of light hadrons \cite{machcone,star_new}, 
and (iii) for non-central collisions, anisotropic collective flow of the matter induces an $e$-$h$ correlation not associate with the back-to-back correlation of heavy quarks.

Assuming that the problems (i)-(iii) are not fatal,  we make  a
theoretical estimate of the magnitude of the quenching in the $e$-$h$ correlation.
Here we adopt the same kinematic regions as those in preceding experiment \cite{phenix_em} in counting trigger and associate particles.
We take electrons with transverse momentum 2.0 $\leq p^{(e)}_{T}\leq$ 3.0 GeV/$c$ as trigger particles.  
Shown in Fig.~\ref{e-h} (a) is the $e$-$h$ azimuthal correlation for the charged
 hadrons with 0.5 $\leq p^{(h)}_{T}\leq$ 1.0 GeV/$c$.
All the counted particles are in mid-pseudorapidity $|\eta|\leq 0.35$ and the impact parameter is taken to be $b$ = 3.1 fm as before.
In this case, there arise peaks in both near and away sides:
The near-side peak appears from the decay of a heavy meson into electron and a charged hadron.
As the heavy quark energy loss increases, both the near-side and away-side peaks are getting quenched.

To quantify these quenching, we use the quenching factor $I_{AA}$ as defined in Eq.~(\ref{eq:measure}) with $(\phi_{\rm min}, \phi_{\rm max})=(0.5 \pi, 1.5\pi)$ for the away side and $(\phi_{\rm min}, \phi_{\rm max})=(-0.5 \pi, 0.5 \pi)$ for the near side.
In Fig.~\ref{e-h} (b), near and away-side quenching factors $I_{AA}^{e\mbox{-}h}$ with 0.5 $\leq p^{(h)}_T\leq$ 1.0 GeV/$c$ and 1.0 $\leq p^{(h)}_T\leq$ 1.5 GeV/$c$ are shown.
In the ideal situation where the problems (i)-(iii) can be neglected, the drag coefficient $\gamma = 3$ leads to the near-side and away-side quenching factors as small as 0.35. 
It is thus not entirely hopeless to extract some information about the sQGP from $I_{AA}^{e\mbox{-}h}$ although $I_{AA}^{e\mbox{-}\mu}$ allows us to make much more transparent comparison between 
the theory and experiments.

In summary, we have proposed an azimuthal electron-muon correlation 
 as a new leptonic tool
  to study the energy loss of heavy quarks in the strongly interacting quark-gluon plasma.
By using the Langevin equation of the heavy quarks under the background matter described by hydrodynamics, we have found that the away-side quenching factor $I_{AA}^{e\mbox{-}\mu}$ is 
 quite useful to quantify the energy loss of heavy quarks
especially for leptons with high transverse momenta.
We have also shown that the azimuthal electron-hadron correlation 
 could be valuable for studying the heavy quarks in hot matter, although
 there are various complications due to hadronic interactions.

  To correctly draw the information on the drag parameter
 inside the hot matter at RHIC, it is very important to determine
  the $e$-$\mu$ azimuthal correlations experimentally not only for 
 the  nucleus-nucleus collisions but also for the  proton-proton and proton-(deuteron-)nucleus
   collisions.  
Indeed, our quenching factor in Eq.~(\ref{eq:measure}) is defined as  
the ratio between the correlation in the nucleus-nucleus collisions and that in the proton-proton collisions. Moreover, studies on the  width of the peak in the $e$-$\mu$ azimuthal correlation 
in the proton-proton and proton-nucleus collisions may shed lights on 
 the gluon distribution inside the energetic nuclei.
   The QGP at LHC  will have higher temperature and longer life time than that at RHIC,
  so that the shape and magnitude of $I_{AA}^{e\mbox{-}\mu}$ would be modified 
  even substantially. 

 Finally, we mention possible theoretical  improvements in the future:
 In our Langevin equation, we adopted the simplest form of the drag parameter and isotropic noise
distribution. If the noise is not isotropic and depends on the direction of the heavy quark
 in the rest frame of the fluid as suggested by the gauge/gravity  correspondence 
 \cite{AdS_Drag}, the shape and width of the azimuthal $e$-$\mu$ correlation could be affected. 
 Possible interactions of the $D$ and $B$ mesons with the hadronic matter before the semileptonic decays, 
which is neglected in our calculation, may affect the magnitude of the  away-side quenching.

Y.~Akamatsu is supported by JSPS fellowships for Young Scientists.
T.~Hatsuda is partially supported by No.~2004, Grant-in-Aid for Scientific Research on Innovative Areas.
T.~Hirano is partially supported by Grant-in-Aid for Scientific Research No.~19740130 and by Sumitomo Foundation No.~080734.

\bibliographystyle{apsrev}

\begin{thebibliography}{99}

\bibitem{QGP}
K.~Yagi, T.~Hatsuda and Y.~Miake, $\it Quark$ - $\it Gluon$ $\it Plasma$
(Cambridge Univ. Press, Cambridge, 2005);
Proceedings of $\it Quark \ Matter \ 2008$,
 J.\ Phys.\ G:\ Nucl.\ Part.\ Phys. {\bf 35} (2008).

\bibitem{Hirano:2008hy}
T.~Hirano, N.~van der Kolk and A.~Bilandzic, arXiv:0808.2684 [nucl-th].

\bibitem{Linde:1980ts}
  A.~D.~Linde,  Phys.\ Lett.\  B {\bf 96}, 289 (1980);
  T.~Hatsuda and T.~Kunihiro, Phys.\ Rev.\ Lett.\  {\bf 55}, 158 (1985);
  C.~E.~DeTar,  Phys.\ Rev.\  D {\bf 32}, 276 (1985).

\bibitem{Son:2007vk}
D.~T.~Son and A.~O.~Starinets,
  Ann.\ Rev.\ Nucl.\ Part.\ Sci.\  {\bf 57}, 95 (2007).

\bibitem{Nakamura:2004sy}
  A.~Nakamura and S.~Sakai, Phys.\ Rev.\ Lett.\  {\bf 94}, 072305 (2005);
  H.~B.~Meyer, Phys.\ Rev.\ Lett.\  {\bf 100}, 162001 (2008).

\bibitem{Bjorken:1982tu}
  J.~D.~Bjorken, FERMILAB-PUB-82-059-THY, unpublished (1982).

\bibitem{Gyulassy:1990ye}
  M.~Gyulassy and M.~Plumer,  Phys.\ Lett.\  B {\bf 243}, 432 (1990);
 X.-N.~Wang and M.~Gyulassy, Phys.\ Rev.\ Lett.\  {\bf 68}, 1480 (1992).

\bibitem{Matsui:1986dk}
  T.~Matsui and H.~Satz,  Phys.\ Lett.\  B {\bf 178}, 416 (1986);
  T.~Hashimoto, K.~Hirose, T.~Kanki, and O.~Miyamura,
  Phys.\ Rev.\ Lett.\  {\bf 57}, 2123 (1986).
 
\bibitem{Rapp}
  R.~Rapp and H.~van Hees,  arXiv:0903.1096 [hep-ph].

\bibitem{phenix2007} 
A.~Adare {\it et al.} (PHENIX Collaboration), 
Phys.\ Rev.\ Lett.\ {\bf98}, 172301 (2007).

\bibitem{star2007}
  B.~I.~Abelev {\it et al.}  (STAR Collaboration),
  Phys.\ Rev.\ Lett.\  {\bf 98}, 192301 (2007).

\bibitem{akamatsu}
Y.~Akamatsu, T.~Hatsuda, and T.~Hirano,
Phys.\ Rev.\ C\ {\bf 79}, 054907 (2009).

\bibitem{Moore:2004tg}
  G.~D.~Moore and D.~Teaney, Phys.\ Rev.\  C {\bf 71}, 064904 (2005).

\bibitem{vanHees:2004gq}
  H.~van Hees and R.~Rapp,  Phys.\ Rev.\  C {\bf 71}, 034907 (2005).

\bibitem{Armesto:2005mz}
  N.~Armesto, M.~Cacciari, A.~Dainese, C.~A.~Salgado, and U.~A.~Wiedemann,
  Phys.\ Lett.\  B {\bf 637}, 362 (2006).

\bibitem{Wicks2007}
S.~Wicks, W.~Horowitz, M.~Djordjevic, and M.~Gyulassy, Nucl.\ Phys.\ {\bf A784}, 426 (2007).
 
\bibitem{Gossiaux:2008jv}
  P.~B.~Gossiaux and J.~Aichelin, Phys.\ Rev.\  C {\bf 78}, 014904 (2008).
 
\bibitem{akamatsu_QM09}
 A preliminary account on $e$-$\mu$ correlation was given in, 
 Y.~Akamatsu, T.~Hatsuda, and T.~Hirano,
 talk at \textit{Quark \ Matter \ 2009} (Knoxville, USA, March 29-April 4, 2009).
 
\bibitem{wang}
G.~Wang \textit{et al.} (STAR Collaboration),
J.\ Phys.\ G:\ Nucl.\ Part.\ Phys. {\bf 35}, 104107 (2008).

\bibitem{star_new}
B.~Biritz, talk at \textit{Quark \ Matter \ 2009} (Knoxville, USA, March 29-April 4, 2009).

\bibitem{machcone}
G.~Torrieri, B.~Betz, J.~Noronha, and M.~Gyulassy, arXiv:0901.0230 [nucl-th].
 
\bibitem{Hirano_Tsuda}
  T.~Hirano and K.~Tsuda, Phys.\ Rev.\  C {\bf 66}, 054905 (2002).
 
\bibitem{Debbasch}
F.~Debbasch, K.~Mallick, and J.~P.~Rivet,
J. of Stat. Phys. {\bf 88}, 945 (1997);
F.~Debbasch and J.~P.~Rivet, J. of Stat. Phys. {\bf 90}, 1179 (1998);
C.~Chevalier and F.~Debbasch, J. Math. Phys. {\bf 49}, 043303 (2008).

\bibitem{AdS_Drag}
S.~S.~Gubser, Phys.\ Rev.\ D {\bf 74}, 126005 (2006);
J.~Casalderrey-Solana and D.~Teaney, Phys.\ Rev.\ D {\bf 74}, 085012 (2006);
C.P.~Herzog, A.~Karch, P.~Kovtun, C.~Kozcaz and L.G.~Yaffe
JHEP\ {\bf 07}, 013 (2006).

\bibitem{PYTHIA}
  T.~Sjostrand, S.~Mrenna and P.~Skands, JHEP {\bf 0605}, 026 (2006).

\bibitem{Gubser:2006qh}
  S.~S.~Gubser,  Phys.\ Rev.\  D {\bf 76}, 126003 (2007).

\bibitem{phenix_em}
T.~Englemore, talk at \textit{Quark \ Matter \ 2009} (Knoxville, USA, March 29-April 4, 2009).

\end{thebibliography}

\end{document}